\documentclass[12pt,pra,tightenlines,notitlepage]{revtex4-1} 

\usepackage{amsmath}  % needed for \tfrac, \bmatrix, etc.
\usepackage{amsfonts} % needed for bold Greek, Fraktur, and blackboard bold
\usepackage{graphicx} % needed for figures

\begin{document}

\title{The evolution of piecewise polynomial wave functions}

\author{Mark Andrews}
\email{mark.andrews@anu.edu.au} % optional
\affiliation{Department of Quantum Science, Australian National University, ACT2601, Australia}
\date{\today}

\begin{abstract}
For a non-relativistic particle, we consider the evolution of wave functions that consist of polynomial segments, usually joined smoothly together. These spline wave functions are compact (that is, they are initially zero outside a finite region), but they immediately extend over all available space as they evolve. The simplest splines are the square and triangular wave functions in one dimension, but very complicated splines have been used in physics. In general the evolution of such spline wave functions can be expressed in terms of antiderivatives of the propagator; in the case of a free particle or an oscillator, all the evolutions are expressed exactly in terms of Fresnel integrals. Some extensions of these  methods to two and three dimensions are discussed.
\end{abstract}


\maketitle % title page is now complete

\section{Introduction}

Functions consisting of polynomial segments, usually joined smoothly at their junctions, will be referred to here as {\it splines}. Such splines have been used in a wide variety of contexts: in approximation and interpolation of functions, and in graphical rendering. In physics splines have been widely used to find approximate energy eigenfunctions. [For a survey, see Ref.\,\onlinecite{Bsplines}.] Here we consider how spline wave functions evolve in time for a free particle, and the results extend easily to a harmonic oscillator. In both cases, the exact evolved wave function can be expressed in terms of Fresnel integrals. Simple examples of such spline wave functions in one dimension are shown in Fig.\,1. We also consider the evolution of splines in 2D and in 3D. In principle, the method can be applied to evolution under other forces, but none are considered here except centrifugal forces.

Normalisable spline wave functions must be zero outside a finite region, but they will immediately extend over all space as they freely evolve. Strictly such wave functions represent a particle initially confined to a finite region and then released from the confining forces, but in practice they could be applied usefully to a wider range of situations. 

There are very few wave functions with compact support for which the exact evolution is known, either for a free particle or for a harmonic oscillator. One known case is the square wave function\,\cite{Anal}, where the evolution has been expressed in terms of error functions with complex argument (which can also be expressed in terms of real Fresnel integrals), using the integral over the propagator; but this wave function has infinite mean energy. The simple procedure discussed here yields the evolution not only of the square wave function but also a wide class of more physically realistic wave functions. Since splines are used to derive approximations to wave functions that arise in a variety of physical situations, the method used here could enable practical procedures to calculate the subsequent evolution when the forces change.

\section{Splines in one dimension}

 We take a spline of degree zero to be a discontinuous function consisting of constant segments; for example, a square wave. A {\textit {linear spline}} (degree one) is continuous and each segment is a straight line; for example, a triangular wave. A second-degree or {\textit {quadratic spline}} is differentiable and made of quadratic segments. Some simple examples of splines of degree 0, 1 and 2 are shown in Fig.\,1.
\begin{figure}[h!]
\centering
\includegraphics[width=10cm]{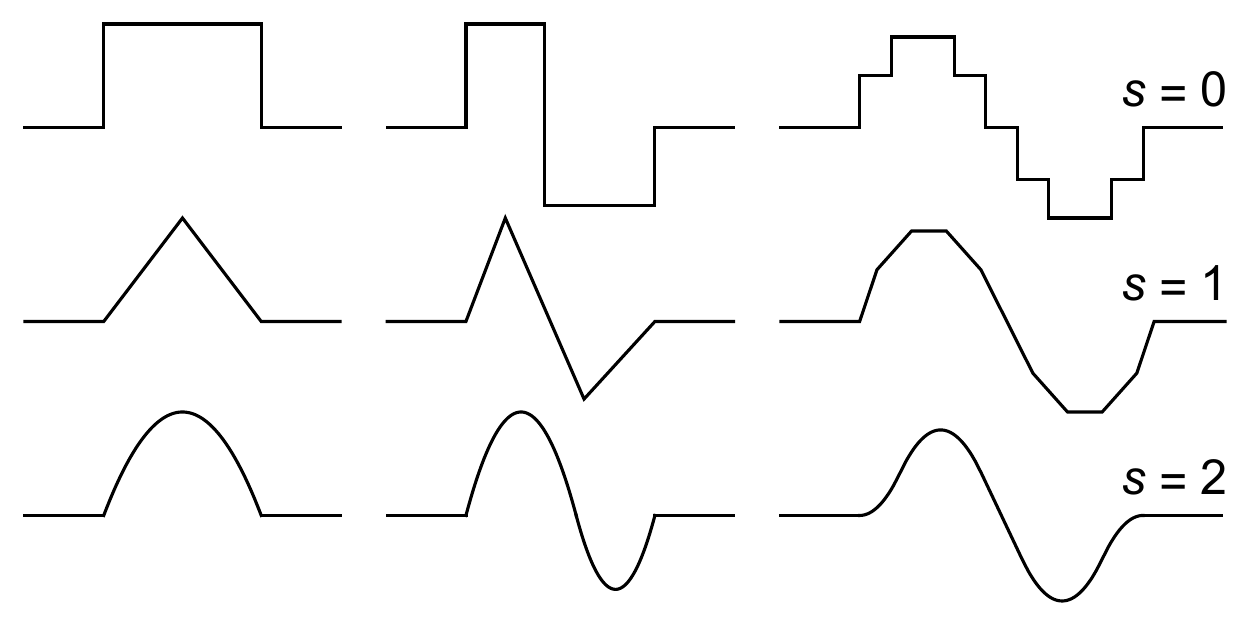}
\caption{Some example of simple spline wave functions of degrees 0, 1 and 2.}
\label{fig:3degrees}
\end{figure}

Thus a 1D spline consists of $n-1$ polynomial segments meeting at the junction points $x=a_i$ with $i=1, 2,...,n$ and $a_1<a_2<...<a_n$. Except for splines of degree zero, a spline $\phi(x)$ is continuous at the junction points and we write $\phi_i\equiv\phi(a_i)$. The simplest class of normalisable splines has the same degree of differentiability at all junctions, including the end points, and we will refer to these as \textit{pure} splines. Only the $s^{\textrm{th}}$ derivative of $\phi(x)$ can jump at any junction, including the end points. Thus all derivatives other than the $s^{\textrm{th}}$ must be zero in the first and last segments. In the first segment, $\phi(x)=(x-a_1)^s\phi_2\,/(a_2-a_1)^s$ for $a_1<x<a_2$. A {\textit {regular spline}} is pure except at the end points.

\vspace{2mm}
\noindent\textbf{Constructing a spline} through the set $\phi_j$. Using the notation $\phi^{(j)}(x)\equiv\!d^j\!\phi(x)/dx^j$, we require that $\phi^{(s-1)}(x)$ is continuous at all internal points. Then we can construct the spline from the $\phi_i$ and the set of derivatives at one end (or any other point) using the Taylor expansions:
\vspace{-4mm}
\begin{eqnarray}
\phi_{j+1}&=&\phi_j+\sum\nolimits_{i=1}^s d_{j+1}^{\,\,i} \phi_j^{(i)}/i!\label{eqn:p0}\\
\phi_{j+1}^{(k)}&=&\phi_j^{(k)}+\sum\nolimits_{i=1}^{s-k}d_{j+1}^{\,\,i} \phi_j^{(k+i)}/i!,\label{eqn:pj}
\end{eqnarray}
where $d_{j+1}=a_{j+1}-a_j$. If we know $\phi_j$ and $\phi_{j+1}$ and all $\phi_j^{(k)}$ except $\phi_j^{(s)}$, then we can calculate $\phi_j^{(s)}$ and all $\phi_{j+1}^{(k)}$ except $\phi_{j+1}^{(s)}$. Thus we can go through all segments to the end point.

For example, in the case of a quadratic spline, Eqns.~(\ref{eqn:p0}) and (\ref{eqn:pj}) with $s=2$ give 
\begin{eqnarray}\label{eqn:step'}
\phi'_{j+1}&=&2d_{j+1}^{-1}(\phi_{j+1}-\phi_{j})-\phi'_{j}\\
\phi''_{j+1}&=&d_{j+1}^{-1}(\phi'_{j+1}-\phi'_{j}).
\end{eqnarray}
[Note that $\phi''(x)$ is constant within each segment, and $\phi''_{j}$ is taken to be $\phi''(x)$ in the segment with $a_{j-1}<x<a_{j}$.] The set $\phi_i$ and one particular $\phi_k'$ determine the spline. If the spline is to represent a function with a distinct maximum, one might choose to force $\phi'$ to be zero there and this would determine the spline (which may or may not be pure).

\vspace{2mm}
\noindent\textbf{Evolvable form of a pure spline}. 
We now show how any pure spline can be expressed in a form that enables its evolution its evolution in terms of antiderivatives of the propagator. The derivative of a pure spline of degree $j$ is a pure spline of degree $j\!-\!1$ and therefore the $(s+1)^{\textrm{th}}$ derivative of a pure spline of degree $s$ is a sum of $\delta$-functions at the junction points $x\!=\!a_i$, including the end points. Thus
\begin{equation}\label{eqn:phis_h}
\phi^{(s+1)}(x)=\sum_{i=1}^n h_i \delta(x\!-\!a_i),
\end{equation}
where the $h_i$ may be complex. The functions
\begin{equation}\label{fs}
f_j(x)=|x|x^{j-1}/2j!
\end{equation}
are successive antiderivatives of $\delta(x)$, with $j=0, 1, 2,...$; these functions have $f'_j(x)\!=\!f_{j-1}(x)$ and $f'_0(x)\!=\!\delta(x)$. Integrating Eq.(\ref{eqn:phis_h}) gives $\phi^{(s)}(x)=\sum_{i=1}^n h_i f_0(x\!-\!a_i)+c_0$ where the $c_0$ is a constant of integration. Then for $x>a_n$ we have $\phi^{(s)}(x)=\frac{1}{2}\sum_{i=1}^n h_i+c_0$, and for $x<a_1$ we have $\phi^{(s)}(x)=-\frac{1}{2}\sum_{i=1}^n h_i+c_0$. Since $\phi^{(s)}(x)$ must be zero in both cases, we require $c_0=0$ and $\sum_{i=1}^n h_i=0$. 

Now integrating $\phi^{(s)}(x)=\sum_{i=1}^n h_i f_0(x\!-\!a_i)$ gives $\phi^{(s-1)}(x)=\sum_{i=1}^n h_i f_1(x\!-\!a_i)+c_1$. Then for $x>a_n$, $\phi^{(s-1)}(x)=\frac{1}{2}\sum_{i=1}^n h_i(x-a_i)+c_1=-\frac{1}{2}\sum_{i=1}^n h_ia_i+c_1$, and for $x\!<\!a_1$, $\phi^{(s-1)}(x)=-\frac{1}{2}\sum_{i=1}^n h_i(x-a_i)+c_1=\frac{1}{2}\sum_{i=1}^n h_ia_i+c_1$. Since $\phi^{(s-1)}(x)$ must be zero in both cases, we require $c_1=0$ and $\sum_{i=1}^n h_ia_i=0$.

This procedure can be continued and the result is that any normalisable pure spline of degree $s$ can be expressed in the form 
\begin{equation}\label{eqn:phi_h}
\phi(x)=\sum\nolimits_{i=1}^n h_i f_s(x\!-\!a_i),
\end{equation}
where $h_i$ is the jump in $\phi^{(s)}(x)$ at $x=a_i$, i.e. $h_i=\phi^{(s)}(a_i +\epsilon)-\phi^{(s)}(a_i -\epsilon)$, and the $h_i$ will satisfy\begin{equation}\label{eqn:hai}
\hspace{10mm}\sum\nolimits_{i=1}^n h_i a_i^j=0\:\:\:\textrm{for}\:\:j=0, 1, 2,...,s.
\end{equation}

Thus, given the set of $\phi_j$ for a pure spline, we can calculate all $\phi_j^{(k)}$ from Eq.(\ref{eqn:pj}). Then $h_j=\phi_{j+1}^{(s)}-\phi_{j}^{(s)}$ and the evolvable form of the spline is given by Eq.(\ref{eqn:phi_h}).

Splines can be used in different ways and the method of construction will vary. There may be more efficient methods than using Eqs.~(\ref{eqn:p0}) and (\ref{eqn:pj}). For example, \[\phi_j=\sum\nolimits_{i=1}^{j-1}h_i f_s(a_j-a_i)-\sum\nolimits_{i=j+1}^n h_i f_s(a_j-a_i)\] and therefore, using Eq.(\ref{eqn:hai}), 
\begin{equation}\label{eqn:phi-h}
\phi_j=2\sum\nolimits_{i=1}^{j-1}h_i f_s(a_j-a_i). 
\end{equation}
This equation provides an alternative way to successively calculate the $h_j$ from the set of $\phi_i$.

A pure spline can be defined by its degree $s$, the set of junction coordinates $\mathfrak{a}=(a_1,a_2,..a_n)$ and the set $\mathfrak{h}=(h_1,h_2,..h_n)$, such that $h_j$ gives the change in $\phi_j^{(s)}$ at $x=a_j$. Given a pure spline, all the splines with the same $\mathfrak{a}, \mathfrak{h}$ but lower $s$ (the derivatives of the given spline) will also be pure, but those with with greater $s$ may not be pure.
%Similarly to the derivation of Eq.(\ref{eqn:hhb}), it follows that
%\begin{equation}\label{eqn:hhc}
%\phi'_j = 2\sum\nolimits _{i=1}^{j-1}h_i(a_j-a_i),
%\end{equation}
%which implies $(\phi'_{j+1}-\phi'_j)/d_{j+1}=2\sum_{i=1}^j h_i$ and therefore
%\begin{equation} \label{eqn:dh1}
%h_j=\frac{1}{2}\,[d_j^{-1}\phi'_{j-1}\!-(d_j^{-1}\!+d_{j+1}^{-1})\phi'_{j}\!+d_{j+1}^{-1}\phi'_{j+1}].
%\end{equation}  \textit{i.e.}\! 

\section{Evolution of pure 1D splines}

The evolution of an initial wave function $\phi(x)$ is given by the propagator equation,
\begin{equation}\label{eqn:Prop}
\psi(x,t)=\int_{-\infty}^{\infty}K(x,x',t)\,\phi(x')\,dx'.
\end{equation}
%where\cite{M1} the propagator $K(x,x',t)$ satisfies Schr\"odinger's equation $\imath\hbar\,\partial_t K=\hat H K$, and $K(x,x',t)\to\delta(x-x')$ as $t\to 0$. 
For a spline, integrating %Eq.(\ref{eqn:Prop}) 
this equation $s$ times by parts,
\begin{equation}\label{eqn:dProp}
\psi(x,t)=(-1)^s\int_{-\infty}^{\infty}K_s(x,x',t)\,\phi^{(s)}(x')\,dx',
\end{equation}
where $K_j(x,x',t)=\partial_{x'}K_{j+1}(x,x',t)$; that is, $K_{j+1}(x,x',t)$ is an antiderivative (or indefinite integral) of $K_j(x,x',t)$ with respect to $x'$, and $K_0(x,x',t)$ is an antiderivative of $K(x,x',t)$. Then, since $\phi^{(s+1)}(x)\!=\!\mbox{$\sum_{i=1}^n h_i \delta(x\!-\!a_i)$}$, and deleting any constants of integration as in Eq.(\ref{eqn:phi_h}),
\begin{equation}\label{eqn:psiKs}
\psi(x,t)=(-1)^{s+1}\sum\nolimits_{i=1}^n h_i K_{s+1}(x,a_i,t).
\end{equation}

\section{Free evolution of 1D splines}

\begin{figure}[h!]
\centering
\includegraphics[width=10cm]{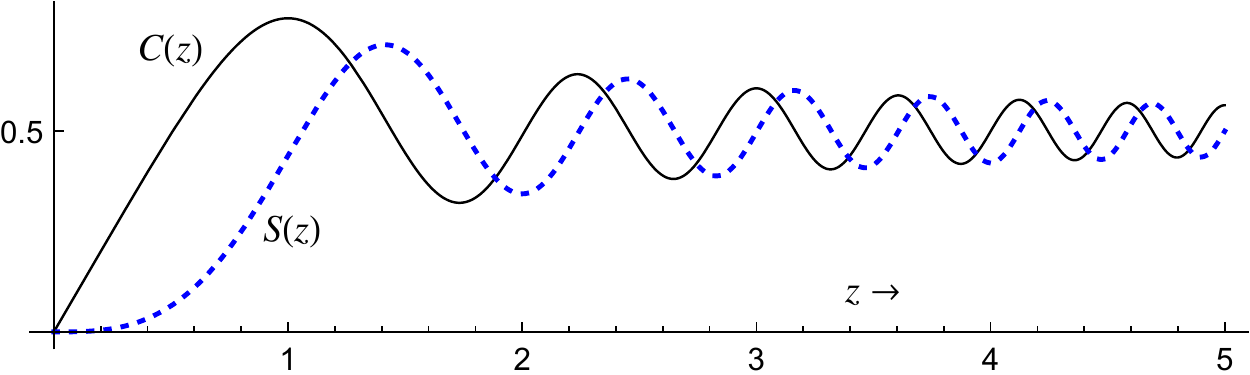}
\caption{The Fresnel integrals $C(z)$ and $S(z)$.}
\label{fig:Fresnel}
\end{figure}

The free propagator\,\cite{M1} is $K(x,x',t)\!=\!\chi(x-x',t)$, where
\begin{equation}\label{eqn:chi}
\chi(x,t):=\sqrt{\frac{m}{2\pi\imath\hbar t}}\exp[\frac{\imath\,m\,x^2}{2\hbar t}].
\end{equation}
The integral over $\chi(x,t)$ can be expressed in terms of the Fresnel integrals\,\cite{F} $C(z)$ and $S(z)$ defined by
\begin{equation}\label{eqn:CS}
\mathcal{E}(z):= C(z)+\imath\,S(z)=\int_0^z\!\exp(\frac{1}{2}\imath\pi u^2)\,du.
\end{equation}
Each of $C(z)$ and $S(z)$ is antisymmetric, real if $z$ is real, and approaches $\frac{1}{2}$ as $z\to\infty$, as shown in Fig.\,\ref{fig:Fresnel}. Hence $\mathcal{E}(z)\sim \sqrt{\imath/2}$ as $z\to\infty$. In terms of the error function (with complex argument), $\mathcal{E}(z)=\sqrt{\imath/2}\,\,\textrm{erf}(\sqrt{\pi/2\imath}\,z)$. 

From Eq.(\ref{eqn:CS}), it follows that
\begin{equation}
\chi_0(x,t):=\int_0^x\!\chi(x',t)\,dx'=\frac{1}{\sqrt{2\imath}}\mathcal{E}(\sqrt{m/\pi\hbar\,t}\,\,x)
\end{equation}
satisfies Schr\"odinger's equation for free evolution
\begin{equation}\label{eqn:SchFree}
\imath\hbar\frac{\partial}{\partial t}\psi(x,t)=-\frac{\hbar^2}{2m}\frac{\partial^2}{\partial x^2}\psi(x,t).
\end{equation}
From the asymptotic behaviour and antisymmetry of the Fresnel integrals, $\chi_0(x,t)\sim \frac{1}{2}|x|/x=f_0(x)$ as $t\to 0$. 

These results can be easily understood qualitatively: as $t\to 0$, $K(x,x',t)\sim \delta(x-x')$ and then the indefinite integral of the $\delta$-function gives the antisymmetric step function and $\chi_0(x,t)$ is its free evolution.

\begin{figure}[h!]
\centering
\includegraphics[width=10cm]{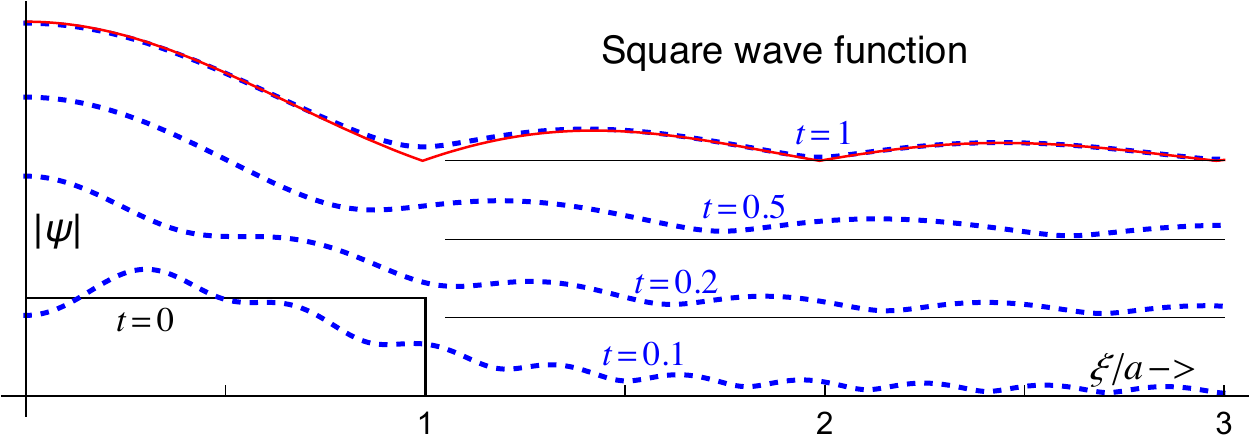}
\caption{The evolution of the square wave function: $|\psi|$ is shown. The lower solid line is for the initial time. The next three dashed curves are for $t=0.1$, $t=0.2$ and $t=0.5$, where the time is in units of $m\,a^2\!/\hbar$; but each curve is raised for separation. The upper two curves overlap considerably and show the evolution at time $t=1$ (dotted) and the asymptotic form for that time (solid) as calculated from the Fourier transform of the  initial wave function. To allow for the spreading of the wave functions, each curve is shown with a different length scale: $|\psi|$ is plotted against $\xi=x\,(1+t^2/\tau^2)^{1/2}$ with $\tau=1/3$. After time $t\!=\!1$ the wave function becomes indistinguishable from the asymptotic form.}
\label{fig:sq3}
\end{figure}

\vspace{2mm}
\noindent\textbf{Example: The square wave function}, $\psi_{sq}(x,0)=1$ for $|x|<a$ and zero for $|x|>a$, is the simplest example of a spline wave function of degree zero. Its derivative is $\partial_x\psi_{sq}(x,0)=\delta(x+a)-\delta(x-a)$, so that $h_1=1$ and $h_2=-1$. Therefore its free evolution is
\begin{equation}\label{eqn:sq}
\psi_{sq}(x,t):=\chi_0(x+a,t)-\chi_0(x-a,t)
\end{equation}
as shown in Fig.(\ref{fig:sq3}). This result has been found by direct integration over the propagator.\cite{Anal}

\vspace{2mm}
\noindent\textbf{The asymptotic form of free evolution} as $t\to \infty$ is determined by the Fourier transform of the initial wave function.\cite{A} Thus
\begin{equation}\label{eqn:asym}
\psi(x,t)\sim \sqrt{m/\imath\hbar t}\,\exp(\imath m X^2/2\hbar t)\,\Phi(m X/\hbar t),
\end{equation}
where $X:=x-\langle\hat x\rangle$ and $\Phi(k)$ is the Fourier transform
\begin{equation}\label{eqn:FT}
\Phi(k)=\frac{1}{\sqrt{2\pi}}\int_{-\infty}^{\infty}\exp(-\imath k x)\psi(x,0) dx.
\end{equation}
Usually, $\psi(x,t)$ will be well approximated by this asymptotic form for all times beyond about $ma^2/\hbar$, where $a$ gives the scale of the initial wave function.

The square wave function with $\psi(x,0)=1$ for $|x|<a$, has the Fourier transform $\sqrt{2/\pi}\,\sin(ka)/k$. Therefore
\begin{equation}
\psi(x,t)\sim \sqrt{\frac{2}{\imath\pi\sigma}}\,\exp\big(\frac{\imath\,x^2}{2a^2\sigma}\big)\frac{\sin(x/a\sigma)}{x/a\sigma},
\end{equation}
where $\sigma=\hbar t/ma^2$, also shown in Fig.\,(\ref{fig:sq3}).

\vspace{2mm}
\noindent\textbf{Free evolution for any degree}. The function
\begin{equation}
\mathcal{E}_1(z):= z\,\mathcal{E}(z)+ \frac{\imath}{\pi}\exp(\frac{1}{2}\imath\pi z^2),
\end{equation}
is an antiderivative of $\mathcal{E}(z)$ with $\mathcal{E}_1(z)\sim\sqrt{\imath/2}\,|z|$ as \mbox{$z\!\to\!\pm\infty$}. If the sequence $\mathcal{E}_n(z)$ is defined by 
\begin{equation}\label{eqn:recurrE_n}
\mathcal{E}_n(z)=\frac{1}{n}\big{[}z\,\mathcal{E}_{n-1}(z)+\frac{\imath}{\pi}\mathcal{E}_{n-2}(z)\big{]}
\end{equation}
for $n>1$ with $\mathcal{E}_0(z)\equiv \mathcal{E}(z)$, then it can be seen that $\mathcal{E}'_n(z)=\mathcal{E}_{n-1}(z)$. From $\mathcal{E}(z)\sim \sqrt{\imath/2}\,|z|/z$ as $|z|\to\pm\infty$, we have $\mathcal{E}_n(z)\sim \sqrt{\imath/2}\,|z|z^{n-1}/n!$. Furthermore Eq.(\ref{eqn:recurrE_n}) ensures that the wave function 
\begin{equation}\label{eqn:chi_n}
\chi_n(x,t):= \frac{1}{\sqrt{2\imath}}(\frac{\pi\hbar t}{m})^{n/2}\mathcal{E}_n(\sqrt{\frac{m}{\pi\hbar t}}\,x)
\end{equation}
satisfies Schr\"odinger's Eq.(\ref{eqn:SchFree}) and 
$\chi_n(x,t)\sim f_n(x)=\frac{1}{2}|x|x^{n-1}/n!$ as $t\!\to\!0$.

Thus for any initial wave function that can be expressed as a linear combination of the $f_n(x)$ we have the exact evolution in terms of $\chi_n(x,t)$. %All $\mathcal{E}_n(z)$ can be expressed as a linear combination of $\mathcal{E}(z)$ and $\exp (\frac{1}{2}\imath\pi z^2)$, with coefficients that are polynomial in $z$, and can therefore be readily calculated.

\subsection{Linear splines}
From Eq.(\ref{eqn:chi_n}) the evolution of $f_1(x)$ is
\begin{equation}
\chi_1(x,t)\!:=\!\frac{1}{\sqrt{2\imath}}[x\,\mathcal{E}(\sqrt{\frac{m}{\pi\hbar t}}\,x)+\imath\sqrt{\frac{\hbar t}{\pi m}}\exp(\frac{\imath mx^2}{2\hbar t})].
\end{equation}
A linear spline that is zero at the two end points is pure and its free evolution is $\sum\nolimits_{i=1}^n h_i \chi_1(x\!-\!a_i,t)$. To find the values of $h_j$ given the set $\phi_j=\phi(a_j)$, note that Eq.(\ref{eqn:phi-h}) implies that $(\phi_{j+1}-\phi_j)/d_{j+1}=\sum_{i=1}^j h_i$, where $d_j:=a_{j}-a_{j-1}$, and therefore
\begin{equation} \label{eqn:h1}
h_j=d^{-1}_j\phi_{j-1}-(d^{-1}_j+d^{-1}_{j+1})\phi_{j}+d^{-1}_{j+1}\phi_{j+1}.
\end{equation}
For equal spacing $d$ between the points,
\begin{equation}\label{eqn:h-equal}
h_j=d^{-1}(\phi_{j-1}-2\phi_{j}+\phi_{j+1}).
\end{equation}

\vspace{2mm}
\noindent\textbf{The Fourier transform}, as defined in Eq.(\ref{eqn:FT}),  of $|x|$ is $-\sqrt{2/\pi}\,k^{-2}$ (as a generalised function \cite{L}). Hence the Fourier transform of $\phi(x)=\sum_{i=1}^n h_i f_1(x-a_i)$ is
\begin{equation}\label{FT_lin_1D}
\Phi(k)=\frac{-1}{k^{2}\sqrt{2\pi}}\sum\nolimits_{i=1}^n h_i\exp(\imath\,a_i k),
\end{equation}
and this can be used to find the asymptotic form of its evolution for large times, using Eq.(\ref{eqn:asym}). Note that $\Phi(k)$ is not singular at $k=0$ because, for small $k$,
\begin{equation}
\Phi(k)=\frac{-1}{\sqrt{2\pi}}\big[\frac{1}{k^2}\!\sum_{i=1}^nh_i+\frac{\imath}{k}\!\sum_{i=1}^nh_ia_i-\frac{1}{2}\!\sum_{i=1}^nh_ia_i^2\big]\!+O(k),\nonumber
\end{equation}
and the first two terms are zero, from Eq.(\ref{eqn:hai}). When there is some spatial symmetry in the spline, some of the exponential terms may combine to give sines and cosines.

\begin{figure}[h!]
\centering
\includegraphics[width=10cm]{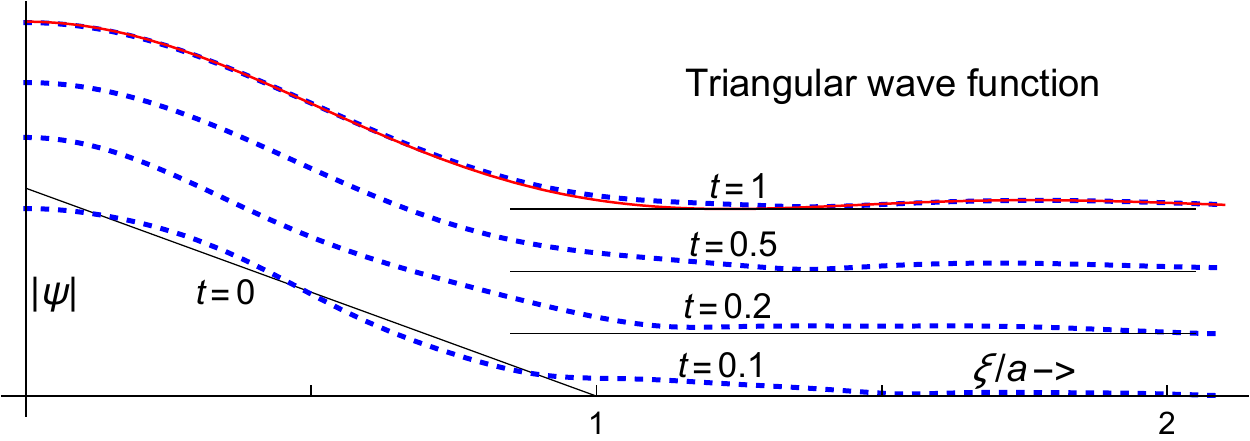}
\caption{The evolution of $|\psi|$ for the symmetrical triangular wave function with base $2a$. The lower solid line is for the initial time and the next three dashed curves are for $t=0.1$, $t=0.2$ and $t=0.5$, where the time is in units of $m\,a^2\!/\hbar$; but each curve is raised for separation. The upper two curves overlap considerably and show the evolution at time $t=1.0$ (dotted) and the asymptotic form for that time (solid) as calculated from the Fourier transform of the  initial wave function. Each curve is shown with a different length scale: $|\psi|$ is plotted against $\xi=x\,(1+t^2/\tau^2)^{1/2}$ with $\tau=1/5$.}
\label{fig:Tri2}
\end{figure}

\vspace{2mm}
\noindent\textbf{The isosceles triangle} is the simplest linear spline. If the length of the base of the triangle is $2a$ then $\mathfrak{a}=(-a,0,a)$. If the height is $1$, then the derivative of the spline $\phi(x)$ is $\phi'(x)=1/a$ for $-a<x<0$ and  $-1/a$ for $0<x<a$. Therefore $\mathfrak{h}=a^{-1}(1,-2,1)$ and $\phi(x)=a^{-1}[f_1(x+a)-2f_1(x)+f_1(x-a)]$ with Fourier transform $\sqrt{2/\pi}\,k^{-2}(1-\cos ka)$. The evolution of this $\phi(x)$ is then $\psi(x,t)=a^{-1}[\chi_1(x+a,t)-2\chi_1(x,t)+\chi_1(x-a,t)]$ 
as shown in Fig(\ref{fig:Tri2}). [A related example is the symmetric trapezium with $\mathfrak{a}\!=\!(-a\!-\!b,-a,a,a\!+\!b)$; if it has unit height then $\mathfrak{h}=b^{-1}(1,-1,-1,1)$.]

\vspace{-6mm}

\subsection{Pure quadratic splines}
From Eq.(\ref{eqn:recurrE_n}) the antiderivative of $\mathcal{E}_1(z)$ is
\begin{equation}
\mathcal{E}_2(z):= \frac{1}{2}z\mathcal{E}_1(z)+\frac{\imath}{2\pi}\mathcal{E}(z)
\end{equation}
and a solution of Schr\"odinger's equation is
\begin{equation}
\chi_2(x,t):=\frac{1}{\sqrt{2\imath}}\frac{\pi\hbar t}{m}\mathcal{E}_2 (\sqrt{\frac{m}{\pi\hbar t}}\,x),
\end{equation}
with $\chi_2(x,t)\to\frac{1}{4}x|x|$ as $t\to 0$. Thus the evolution the quadratic spline $\phi(x)=\sum_{i=1}^n h_i f_2(x-a_i)$ is $\phi(x,t)=\sum\nolimits_{i=1}^n h_i \chi_2(x\!-\!a_i,t)$ and $h_i$ can be calculated from Eq.(\ref{eqn:step'}).

\vspace{2mm}
\noindent\textbf{The Fourier transform} of $x|x|$ is $-\imath\sqrt{8/\pi}\,k^{-3}$, so that the Fourier transform of $\phi(x)=\sum_{i=1}^n h_i f_2(x-a_i)$ is
\begin{equation}
\Phi(k)=-\imath k^{-3}\sqrt{1/2\pi}\sum\nolimits_{i=1}^n h_i\exp(\imath\,a_i k),
\end{equation}
and this can be used to find the asymptotic form of its evolution for large times, using Eq.(\ref{eqn:asym}). Again $\Phi(k)$ is not singular at $k=0$ because, for small $k$,
\begin{equation}
\Phi(k)=\frac{-\imath}{\sqrt{2\pi}}\big[\frac{1}{k^3}\!\sum_{i=1}^nh_i+\frac{\imath}{k^2}\!\sum_{i=1}^nh_ia_i-\frac{1}{2k}\!\sum_{i=1}^nh_ia_i^2\!-\frac{\imath}{6}\!\sum_{i=1}^nh_ia_i^3\!+O(k)\big],
\end{equation}
and the first three terms are zero from Eq.(\ref{eqn:hai}).

\vspace{2mm}
\noindent\textbf{Smooth hump}. The simplest pure quadratic spline is a symmetric hump with $\mathfrak{a}=a(-2,-1,1,2)$ and
\begin{equation} \label{eqn:smoothump}
\psi(x,0)= \left \{ \begin{array}{ll}
 1-x^2/2a^2 & \mbox{if $|x|<a$} \\
 (2a-|x|)^2/2a^2 & \mbox{if $a<|x|<2a$,}
 \end{array}
 \right .
\end{equation}
with $\mathfrak{h}\!=\!a^{-2}(1,-2,2,-1)$ and evolution
$\psi(x,t)\!=\!a^{-2}$ $[\chi_2(x+2a,t)\!-2\chi_2(x+a,t)\!+2\chi_2(x-a,t)\!-\chi_2(x-2a,t)].$ The Fourier transform of the initial wave function is $\sqrt{8/\pi}\,a^{-2}k^{-3}\sin ka\,(1-\cos ka)$.

\subsection{Impure quadratic splines}

If a wave function is initially confined by infinite barriers, its derivative must be finite at the barriers. This requires a multiple of $f_1(x)$ centred on each boundary:
\begin{equation}\label{eqn:phic}
\phi(x)=c_1 f_1(x-a_1)+\sum_{i=1}^n h_i f_2(x-a_i)+c_n f_1(x-a_n).
\end{equation}
For $x>a_n$ we have $f_1(x-a_i)=\frac{1}{2}(x-a_i)$ and $f_2(x-a_i)=\frac{1}{4}(x^2-2a_ix+a_i^2)$. Hence for $\phi(x)\equiv 0$ outside $(a_1,a_n)$
\begin{eqnarray}
&&\sum\nolimits_{i=1}^n h_i = 0,\label{eqn:H0}\\
&&\sum\nolimits_{i=1}^n h_i a_i = c_1+c_n,\label{eqn:H1}\\
&&\sum\nolimits_{i=1}^n h_i a_i^2 = 2(c_1a_1+c_n a_n).\label{eqn:H2}
\end{eqnarray}

Differentiating Eq.(\ref{eqn:phic}) gives
\begin{equation}\label{eqn:dphic}
\phi'(x)\!=\!c_1 f_0(x-a_1)+\sum_{i=1}^n h_i f_1(x-a_i)+c_n f_0(x-a_n), \nonumber
\end{equation}
and the only term that changes suddenly as $x$ crosses $a_1$ is the first term because $f_0$ changes from $-\frac{1}{2}$ to $\frac{1}{2}$. Hence $\phi'_1=c_1$ and similarly $\phi'_n=-c_n$. Thus $c_1$ is the change in the slope of $\phi$ at $x=a_1$, while $h_1$ is the change in the slope of $\phi'$. The free evolution of $\phi(x)$ is $c_1\chi_1(x-a_1,t)+\sum_{i=1}^nh_i\chi_2(x-a_i,t)+c_n\chi_1(x-a_n,t)$.

Given the set $\phi_i$ and the initial slope $c_1$, the constants $h_i$ can be found, as before, from Eq.(\ref{eqn:step'}) starting with $\phi'_1=c_1$ and then $c_n$ will emerge as $-\phi'_n$.

\begin{figure}[h!]
\centering
\includegraphics[width=10cm]{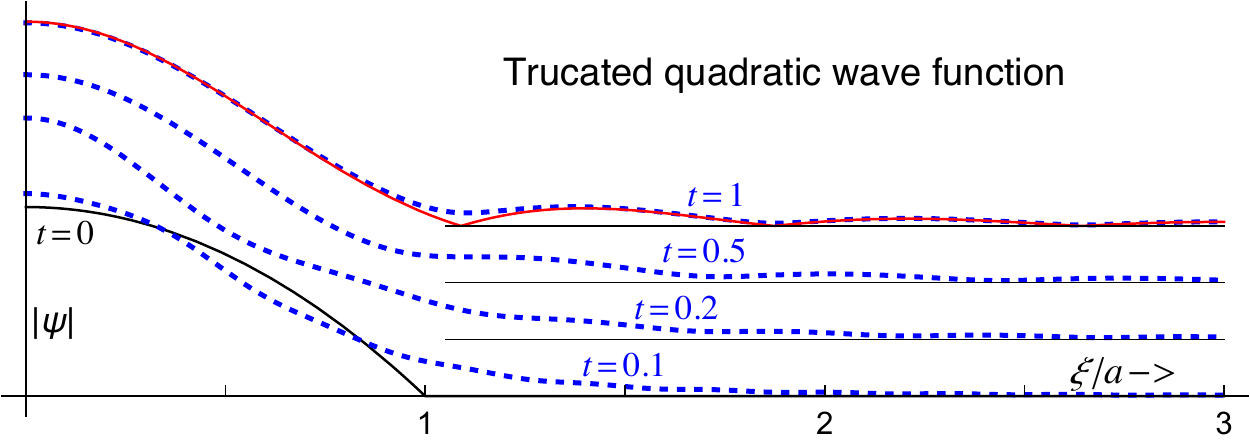}
\caption{The evolution of $|\psi|$ for a truncated quadratic wave function. The lower solid curve is for the initial time. The dashed curves are for $t=0.1$, $t=0.2$ and $t=0.5$ and $t=1.0$, where the time is in units of $m\,a^2\!/\hbar$; but after the first, each curve is raised. The upper solid curve shows the  asymptotic form, also at time $t=1.0$. Each curve is shown with a different length scale, with $\xi=x\,(1+t^2/\tau^2)^{1/2}$ with $\tau\!=\!1/4$.}
\label{fig:Trunc2}
\end{figure}

\vspace{2mm}
\noindent\textbf{The truncated quadratic} is a simple example with
\begin{equation} \label{eqn:truncquad}
\phi(x)=(1-x^2/a^2)\; \mbox{   for $|x|<a$} 
\end{equation}
or $\phi(x)\!=\!$ $2a^{-2} [af_1(x+a)-\!f_2(x+a)+f_2(x-a)+a f_1(x-a)]$. Its Fourier transform is $\sqrt{8/\pi}(\sin ka-ka\cos ka)/a^2k^3$ and its free evolution is $2a^{-2} [a\chi_1(x+a,t)-\chi_2(x+a,t)+\chi_2(x-a,t)+a\chi_1(x-a,t)]$, as shown in Fig.\,(\ref{fig:Trunc2}).

%\begin{equation} \label{eqn:truncquad}
%\phi(x)= \left \{ \begin{array}{ll}
% 1-x^2/a^2 & \mbox{if $|x|<a$} \\
%0 & \mbox{if $|x|\ge a$,}
%\end{array}
% \right .
%\end{equation} 

%and $\phi(x)$ equal to zero otherwise.
%$2a^{-2} [af_1(x\!+\!a)\!-\!f_2(x\!+\!a)\!+\!f_2(x\!-\!a)\!+\!a f_1(x\!-\!a)]$

\section{Splines in two dimensions}
 
\begin{figure}[h!]
\centering
\includegraphics[width=10cm]{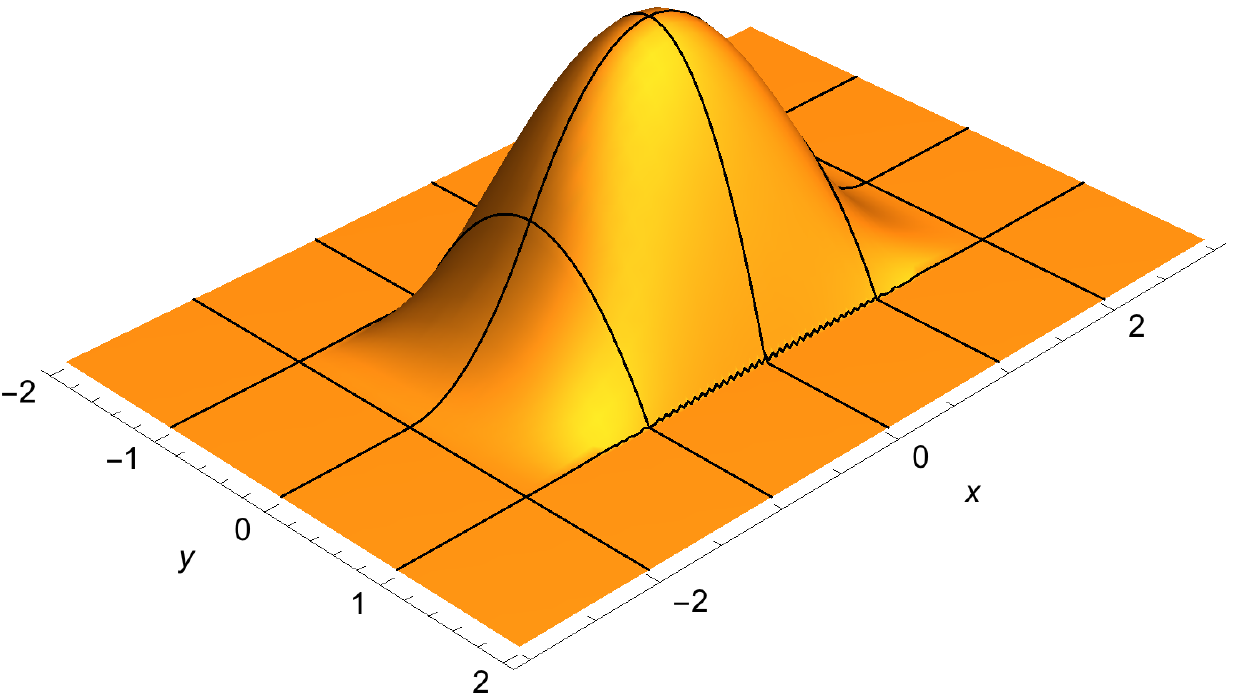}
\caption{The product of a pure quadratic hump $\phi_1(x)$ and an impure truncated quadratic $\phi_2(y)$.}
\label{fig:Product2D}
\end{figure}

A simple generalisation to two dimensions is given by the product $\phi_1(x)\phi_2(y)$ of two 1D splines. For example, if $\phi_1(x)$ is the smooth hump in Eq.(\ref{eqn:smoothump}) and $\phi_2(y)$ is the truncated quadratic in Eq.(\ref{eqn:truncquad}), then the product is the biquadratic spline shown in Fig.\,(\ref{fig:Product2D}). Its evolution is the product of the evolutions given for $\phi_1(x)$ and $\phi_2(y)$. Clearly, we need to consider more general wave functions.

 \vspace{2mm}
\noindent\textbf{Bilinear splines} generated from a given set of complex values $\phi(i,j)$ can be related to a set of 1D linear splines parallel to the $x$-axis and another set parallel to the $y$-axis, intersecting at all the junction points. The set $\phi(i,j)$ completely defines the bilinear spline 
\begin{equation}\label{eqn:phi_lin}
\phi(x,y)=\sum\nolimits_{i,j}h_{ij}f_1(x-a_i)\,f_1(y-b_j).
\end{equation}
As shown in the Appendix \ref{app:find_h}, $h(i,j)$ is determined by $\phi(i,j)$ and its eight nearest neighbours:
\begin{eqnarray}
d^{-2}\,h_{ij}
&=&+\phi_{i-1,j+1}-2\,\phi_{i,j+1}+\phi_{i+1,j+1}\nonumber\\
& &\!\!-2\,\phi_{i-1,j}\;\,+4\,\phi_{i,j}\;\;-2\,\phi_{i+1,j}\nonumber\\
& &+\phi_{i-1,j-1}-2\,\phi_{i,j-1}+\phi_{i+1,j-1}.
\end{eqnarray}
This can be applied to any junction (even on the boundary). Then the evolution is easily calculated by replacing $f_1(u)$ by $\chi_1(u,t)$ in Eq.(\ref{eqn:phi_lin}).

Except for wave functions with rectangular boundaries, there will be artifacts that arise through approximating the boundary by one that lies on the grid. Also the boundary is constrained to not allow any segment to have three sides on the boundary, because the only bilinear function that is zero on three sides is zero throughout.

\begin{figure}[h!]
\centering
\includegraphics[width=10cm]{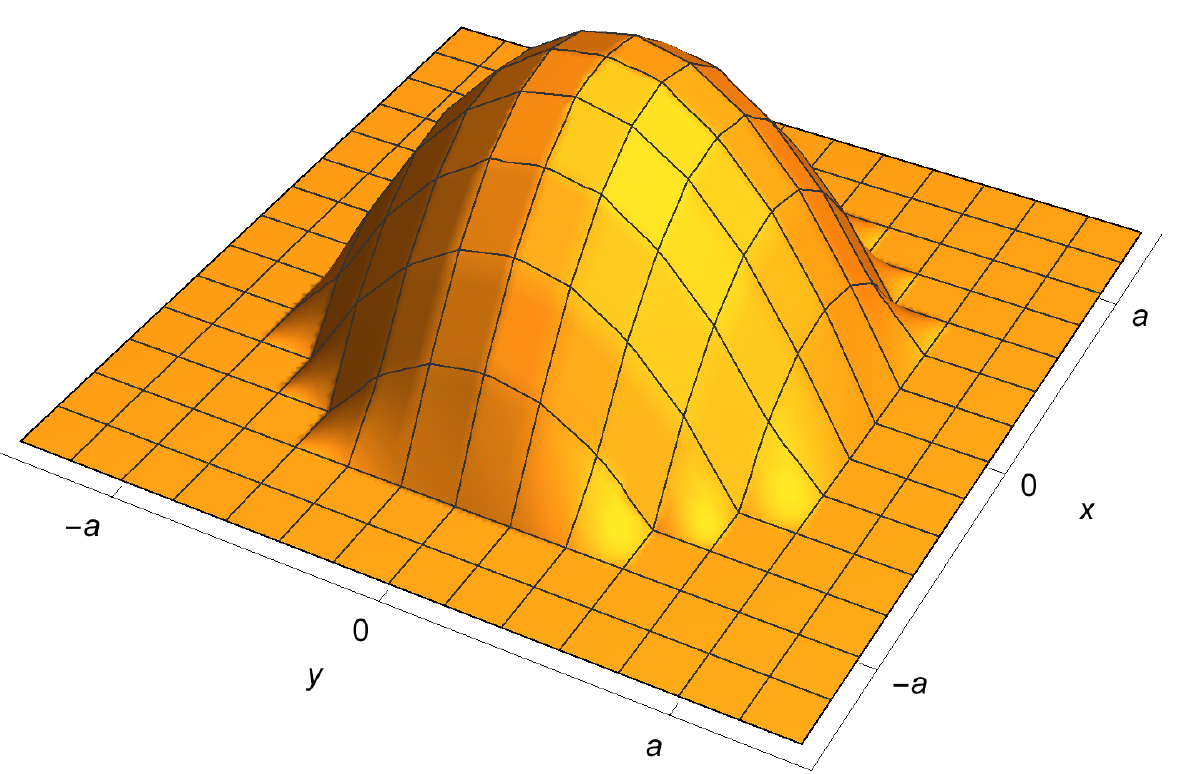}
\caption{The bilinear spline with a grid spacing of $a/5$ that contains the axially symmetric truncated quadratic $\phi(r)=1-r^2/a^2$ for $r<a$.}
\label{fig:Trunc2D}
\end{figure}

 \vspace{2mm}
\noindent\textbf{Example}: the bilinear spline generated by the axially symmetric truncated quadratic $\phi(r)=1-r^2/a^2$ for $r<a$ with grid-spacing $a/5$ is shown in Fig.\,\ref{fig:Trunc2D}, which also shows the effects that come from boundary approximation. 

In three dimensions it is well known that the free evolution of spherically symmetric wave functions $\psi(r)$ can be reduced to the free evolution in one dimension of $\psi(r)/r$. Examples will be given in Section \ref{sec:3D}. For two dimensions the equivalent reduction introduces an additional term, proportional to $r^{-2}$, in the Hamiltonian and the method used above cannot be easily applied. This means that circular boundaries are not easily dealt with.

\begin{figure}[h!]
\centering
\includegraphics[width=165mm]{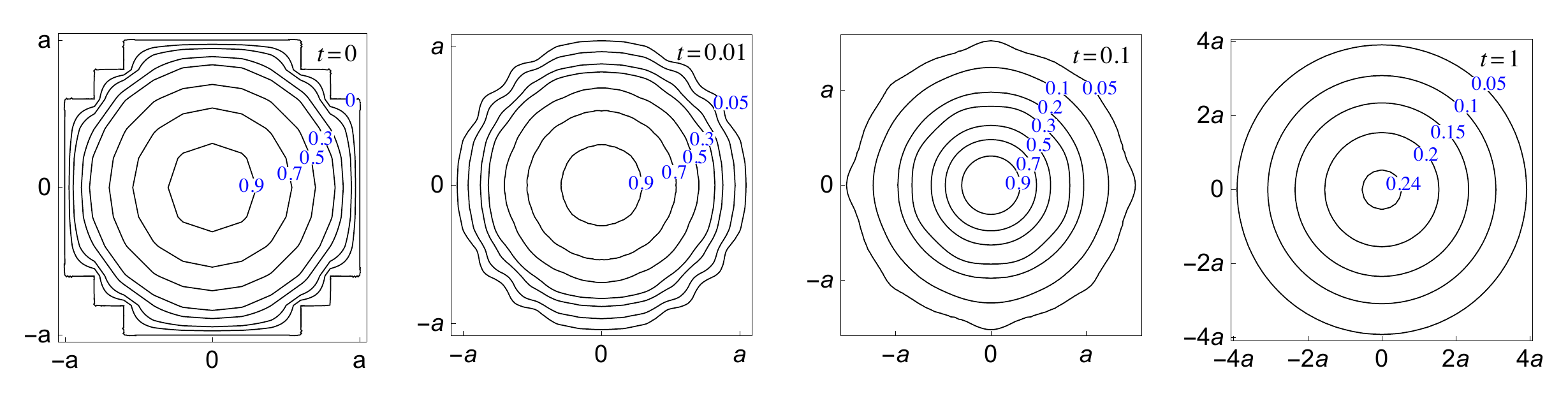}
\caption{Contours of the evolution of the bilinear spline shown in Fig.\,\ref{fig:Trunc2D} at four times, $t=0, 0.01, 0.1, 1$. The numbers on the contours give $|\psi(x,y,t)|$, where the initial wave function has $\psi(0,0,0)=1$. The $x$ and $y$ axes are interchangeable. Note the changes in scale as the wave function spreads. Beyond $t=1$ the contours are almost circular and agree closely with the asymptotic value from Eq.(\ref{eqn:asympBiLin}).}
\label{fig:ContoursT2D}
\end{figure}

 \vspace{2mm}
\noindent\textbf{The evolution of a bilinear spline} $\phi(x,y)$ in the form given in Eq.(\ref{eqn:phi_lin}) is then
\begin{equation}
\psi(x,y,t)=\sum\nolimits_{i,j}h_{ij}\,\chi_1(x-a_i,t)\,\chi_1(y-b_j,t).
\end{equation}
As in the derivation of Eq.(\ref{FT_lin_1D}), the Fourier transform of $\phi(x,y)$ is
\begin{equation}
\Phi(k_x,k_y)=(2\pi)^{-1} (k_x k_y)^{-2} \sum\nolimits_{i,j}h_{i,j}\exp(\imath\,a_i k_x + \imath\,b_j k_y)
\end{equation}
and the asymptotic form of the evolved wave function is
\begin{equation}\label{eqn:asympBiLin}
\psi(x,y,t)\sim \frac{m}{\imath\hbar t}\exp\big[\frac{\imath m (x^2+y^2)}{2\hbar t}\big]\,\Phi(\frac{m x}{\hbar t},\frac{m y}{\hbar t}),
\end{equation}
where we have assumed that the wave function is centered so that $\langle \hat x \rangle=0$ and $\langle \hat y \rangle=0$. As an example the evolution of the bilinear spline in Fig.\,\ref{fig:Trunc2D} is shown in Fig.\,\ref{fig:ContoursT2D}.

\vspace{2mm}
\noindent\textbf{Biquadratic splines}. If we have a set of (possibly complex) values $\phi_{i,j}$ at the junction points (all on a square grid) then we can generate a set of 1D splines joining the points along the $x$-axis and another set for the $y$-axis, as described in Section II, provided that we have also specified the first derivative at one point on each spline. Thus, at each junction point, we have the values of $\phi, \phi_x$ and $\phi_y$. This set of eight independent quantities for each cell does not, in general, completely determine the 2D spline in the cell; there are nine independent constants required to specify a biquadratic spline in a cell, such as $\phi$, $\phi_x$, $\phi_y$, $\phi_{xx}$, $\phi_{xy}$, $\phi_{yy}$, $\phi_{xxy}$, $\phi_{xyy}$, $\phi_{xxyy}$ at one junction. After one cell is specified, all the cells in the spline are determined (although it is not always possible to create a regular 2D spline containing a given set of 1D splines). One way to carry the information required from one cell to its neighbours is via $\phi_{xy}$, which does not change in crossing from one cell to its neighbour at a junction.

For a pure biquadratic spline (zero first derivative normal to every boundary) the spline can be expressed as
\begin{equation}\label{eqn:phi_biQ}
\phi(x,y)=\sum\nolimits_{i,j}h_{ij}f_2(x-a_i)\,f_2(y-b_j)
\end{equation}
and the free evolution follows. When a biquadratic spline is impure (with non zero first derivative normal to some boundary) we must add terms from the discontinuities in the derivatives at the boundary points. The detail on this is given in the Appendix \ref{app:find_h}.

\vspace{2mm}
\noindent\textbf{An example of a biquadratic spline} is shown in Fig.\,{\ref{fig:BiQ}. It has been chosen to show how the discontinuities in the first derivative on some boundaries can be dealt with, in the process of determining the evolution of the spline as a free wave function. It also shows that a corner indent in the boundary can be accommodated if the first derivatives on this part of the boundary are zero (although a similar function with non-zero first derivative there cannot be constructed). 

The spline sits on a 4x4 grid with one corner cell missing and the origin is taken to be at the inner vertex of this corner cell. [The grid is shown in Figs.\ref{fig:data1} and \ref{fig:data2} in the Appendix \ref{app:find_h}.] The spline is symmetric about a diagonal, so that $\phi(x,y)=\phi(y,x)$. Then there are only two non-zero values of the function at grid intersection points: $\phi(1,0)=\phi(0,1)$ and $\phi(1,1)$. We take $\phi(1,0)=\phi(0,1)=1$, which sets the scale of $|\psi|$. We also take the first derivatives of the 1D splines to be zero on the boundaries $x=-1$, $y=-1$ and on the boundaries of the corner indent. With these assumptions, the 2D spline is then uniquely determined by the value of $\phi(1,1)$. The simplest choice is to have no discontinuity in the second derivatives of the 1D splines at that point and this requires $\phi(1,1)=7/4$. Then the 1D spline through $(0,0), (1,0)$ and $(2,0)$ is 
\begin{equation} \label{eqn:Q_1D_1}
\phi_1(x)= \left \{ \begin{array}{ll}
 \,\, x^2 & \mbox{if $0<x<1$} \\
 (2-x)(3x-2) & \mbox{if $\:\:1<x<2$.}
 \end{array}
 \right .
\end{equation}
and the 1D spline from $(-1,1)$ to $(2,1)$ is
\begin{equation} \label{eqn:Q_1D_2}
\phi_2(x)= \left \{ \begin{array}{ll}
 (x+1)^2 & \mbox{if $-1<x<0$} \\
 \frac{1}{4}(2-x)(5x+2) & \mbox{if $\:\:0<x<2$.}
 \end{array}
 \right .
\end{equation}
The complete 2D spline and the details of determining its evolvable form are given in the Appendix \ref{app:find_h}.

\begin{figure}[h!]
\centering
\includegraphics[width=10cm]{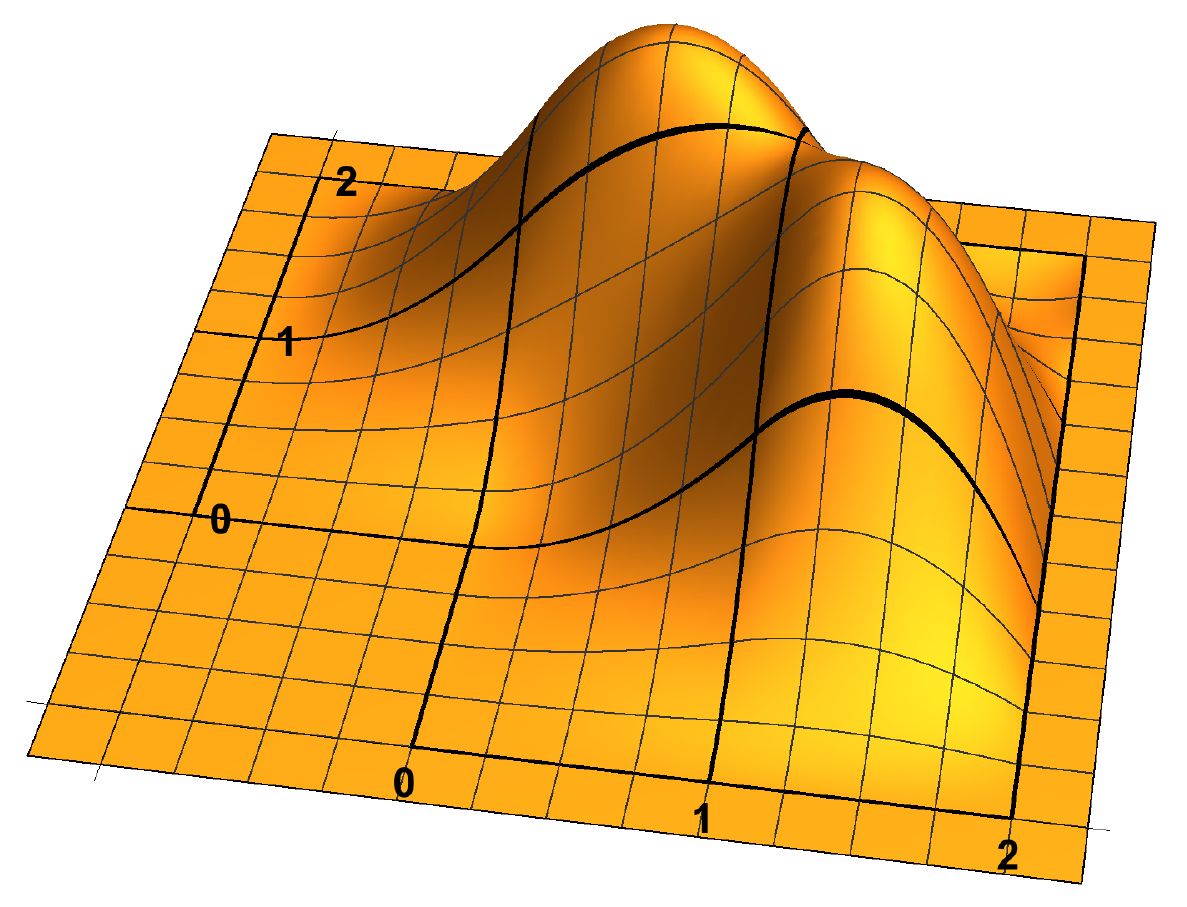}
\caption{An example of an impure biquadratic spline, used to show how to deal with the discontinuities in the first derivatives on some boundaries. This example also shows that a regular biquadratic spline can exist on a grid with a corner cell excised, provided that there are no discontinuities in the first derivatives on the boundaries surrounding that corner. }
\label{fig:BiQ}
\end{figure}

We have dealt only with grids with equal spacing; that is, the junctions lie on a square grid. The extension to unequal spacing is straight forward, but the methods considered here do not easily extend beyond a grid of parallel and perpendicular lines.

\section{Splines in three dimensions}\label{sec:3D}

The rectangular grid used for two dimensions extends simply to three, particularly for trilinear splines. The constraints found for regular biquadratics are more severe for triquadratics, but a spherically symmetric wave function can (apart from a factor of $r$) be reduced to a one-dimensional form, whereas an axially symmetric one in 2D acquires a term in $r^{-2}$ in the potential and this would require a different propagator involving Bessel functions of integer order.

 \vspace{2mm}
\noindent\textbf{A simple example} of a spherically symmetric wave function has the form $\psi(r)=1-r/a$ for $r<a$ and otherwise zero. [This wave function has a discontinuous derivative at $r=0$.] To evolve this wave function, consider first the antisymmetric one-dimensional equivalent, $\phi(x)=x(1-|x|/a)$ which can be expressed as $\phi(x)=\sum_{i=1}^3 h_i f_2(x-a_i)-f_1(x-a_1)+f_1(x-a_3)$ with ${\mathfrak{a}}=(-a,0,a)$ and ${\mathfrak{h}}=2a^{-1}(1,-2,1)$. Then $\psi(r)=\phi(r)/r$ for $r>0$ and its evolution is $\psi(r,t)=r^{-1}[\sum_{i=1}^3 h_i \chi_2(r-a_i)-\chi_1(r-a_1)+\chi_1(r-a_3)]$.

A smoother wave function is $\psi(r)=1-r^2/a^2$ for $r<a$, which corresponds to the 1D cubic $\phi(x)=(1-x^2/a^2)x/a$ for $|x|<a$. This has no internal points, just the two boundary points; but both $\phi'$ and $\phi''$ change at the boundaries. We have $\phi'(\pm a)=-2/a$, $\phi''(\pm a)=\mp2/a^2$, and $\phi'''(\pm a)=-6/a^3$. Therefore the evolvable form is
\begin{equation}
\phi(x)=\sum_{s=1}^3 \sum_{i=1}^2 h_{s,i}\,f_s(x-a_i),
\end{equation}
where ${\mathfrak{a}}=a(-1,1)$ and ${\mathfrak{h_1}}=a^{-1}(-2,2)$, ${\mathfrak{h_2}}=a^{-2}(6,6)$, ${\mathfrak{h_3}}=a^{-3}(-6,6)$. Then the evolution of $\psi(r)$ is
\begin{equation}
\psi(r,t)=r^{-1}\sum_{s=1}^3 \sum_{i=1}^2 h_{s,i}\,\chi_s(r-a_i).
\end{equation}

To extend this approach to wave functions that vary with direction, while still having a spherical boundary, would require the radial Hamiltonian with non-zero angular momentum and this contains a term corresponding to the centrifugal force; this would require a different propagator involving spherical Bessel functions.

\vspace{2mm}
\noindent\textbf{Trilinear splines}. The weights needed to calculate $h_{ijk}$ from the values of $\phi$ at the 27 junctions in the surrounding cube can be guessed from the facts that the weights at the three points on each grid line must have the ratios -1, 2, 1 and each face will then have weight +4 for the centre, -2 for the nearest neighbour, and +1 for the furthest. The simple result for a junction point in 3D is that we need weight 8 at the point, -4 for the 6 nearest neighbours, +2 for the 12 next nearest, and -1 for the 8 furthest points. A more formal proof is given in the Appendix \ref{app:find_h}

\section{Evolution of splines in an oscillator}\label{sec:Osc}

Any free evolution can be easily transformed\,\cite{Takagi,St} into an evolution of the same initial wave function in a harmonic oscillator. Thus, if $\phi(x,t)$ satisfies the free Schr\"odinger Eq.(\ref{eqn:SchFree}) 
%\begin{equation}\label{eqn:SchOsc}
%\imath\hbar\frac{\partial}{\partial t}\psi(x,t)=-\frac{\hbar^2}{2m}\frac{\partial^2}{\partial x^2}\psi(x,t)+\frac{1}{2}m\omega^2 x^2\psi(x,t)
%\end{equation}
we change to new variables $\xi$ and $\tau$, where
\begin{equation}
x=\xi/\cos\omega\tau\hspace{4mm}\textrm{and}\hspace{4mm}t=\tan\omega\tau/\omega,
\end{equation}
and add a phase $\theta(\xi,\tau)=-\frac{1}{2}\tan\omega\tau\,\xi^2/\alpha^2$, where $\alpha=\sqrt{\hbar/m\omega}$ is the intrinsic length scale of the oscillator. The change in the scale of $x$ also requires a factor of $(\cos\omega\tau)^{-1/2}$ to preserve normalisation. Thus, if we insert 
\begin{equation}\label{eqn:trans}
\psi(\xi,\tau):=\frac{\,\,\exp\imath\theta(\xi,\tau)}{\sqrt{\cos\omega\tau}}\,\phi\big(\frac{\xi}{\cos\omega\tau},\frac{\tan\omega\tau}{\omega}\big)
\end{equation}\\
into Schr\"odinger's equation for the oscillator, we obtain
\begin{eqnarray}\label{eqn:toSchOsc}
\big[-\imath\hbar\frac{\partial}{\partial t}-\frac{\hbar^2}{2m}\frac{\partial^2}{\partial \xi^2}+\frac{1}{2}m\omega^2 \xi^2\big]\psi(\xi,\tau)\nonumber =\\
\frac{\exp\imath\theta(\xi,\tau)}{(\cos\omega\tau)^{5/2}}\big[-\imath\hbar\frac{\partial}{\partial t}-\frac{\hbar^2}{2m}\frac{\partial^2}{\partial x^2}\big]\phi(x,t),
\end{eqnarray}
which shows that if $\phi(x,t)$ satisfies the free Schr\"odinger equation then $\psi(x,t)$ satisfies Schr\"odinger's oscillator equation. For example, in the case of the square wave function, with $\psi_{sq}(x,0)=1$ for $|x|<a$, Eq.(\ref{eqn:sq}) leads to
\begin{equation}
\psi_{sq}(x,t)\!=\!\frac{\exp[-\frac{1}{2}\imath\,\tan\omega t\,\,x^2/\alpha^2]}{\sqrt{2\imath\cos\omega t}}[\mathcal{E}(z_+)\!-\!\mathcal{E}(z_-)],
\end{equation}
where $z_\pm=\sqrt{2/(\pi\,\sin 2\omega t)}\,(x\pm a\,\cos\omega t)/\alpha$. This evolution is equivalent to that found in Ref.\,[\onlinecite{Anal}]. The transformation in Eq.(\ref{eqn:trans}) can be applied to any free evolution.

\section{Conclusion}\label{sec:Conc}

We have shown that a wide range of piecewise polynomial wave functions can be simply evolved for a free particle and for an oscillator. Practical applications will usually require further consideration of the most efficient methods of calculation and how best to deal with the ambiguities inherent in non-linear splines in 2D and 3D; but these matters will depend on the application.

We have dealt only with splines that are constructed to pass exactly through a given set of points. In practice, one might allow approximation and this has been studied in the vast spline literature; but whatever method of construction is used, if the result is a piecewise polynomial spline then the exact free (or oscillator) evolution can be found.

There may be systems other than the free particle and the oscillator for which this approach may be useful, but none have been investigated here.

\appendix*

\section{Determining the evolvable form of a spline} \label{app:find_h}

\noindent\textbf{For bilinear splines}, label the four cells surrounding a junction point, taken temporarily to be the origin, by indexes $\alpha = \textrm{sgn}(x)$ and $\beta=\textrm{sgn}
(y)$. Then
\begin{equation}
\phi^{\alpha,\beta}(x,y)=\phi_{00}+c_x^\alpha x+c_y^\beta y+c_{xy}^{\alpha,\beta}x y,
\end{equation}
because $\partial_x\phi$ must be the same immediately above and below the $x$-axis, and similarly for $\partial_y\phi$ on either side of the $y$-axis. Then $\phi_{xy}^{\alpha,\beta}(x,y)=c_{xy}^{\alpha,\beta}$
%\begin{equation}
%\phi_x^{\alpha,\beta}(x,y)=c_x^\alpha+c_{xy}^{\alpha,\beta}y,
%\end{equation}
and
\begin{equation}
\phi_{xy}(x,y)=C+Af_0(x)+Bf_0(y)+Hf_0(x)f_0(y),
\end{equation}
valid over all four cells, where $C=\frac{1}{4}\sum_{\alpha,\beta}c_{xy}^{\alpha,\beta}$, \mbox{$A\!=\!\frac{1}{2}\sum_{\alpha,\beta}\alpha\,c_{xy}^{\alpha,\beta}$,} $B\!=\!\frac{1}{2}\sum_{\alpha,\beta}\beta\,c_{xy}^{\alpha,\beta}$, $H\!=\!\sum_{\alpha,\beta}\alpha\beta\,c_{xy}^{\alpha,\beta}$. Hence $\phi_{xxyy}(x,y)=H\,\delta(x)\delta(y)$. Eq.(\ref{eqn:phi_lin}) gives $\phi_{xxyy}(x,y)=\sum_{i,j}h_{ij}\delta(x-a_i)\,\delta(y-b_j)$ and therefore the value of $h_{ij}$ at any junction equals the value of $H$ calculated from its four neighbouring cells.

The $c_{xy}^{\alpha,\beta}$ can be calculated from the values of $\phi$ at the four corners. In the ++cell (with $x>0$ and $y>0$) we have $c_x^+=\phi_{+0}-\phi_{00}$, $c_y^+=\phi_{0+}-\phi_{00}$ and
\begin{equation}\label{eqn:cxy_lin}
c_{xy}^{++}=+(\phi_{++}+\phi_{00})-(\phi_{+0}+\phi_{0+}).
\end{equation}
Similarly, for the other cells,
\begin{eqnarray}
c_{xy}^{-+} & = & -(\phi_{-+}+\phi_{00})-(\phi_{0+}+\phi_{-0})\\
c_{xy}^{--} & = & +(\phi_{--}+\phi_{00})-(\phi_{-0}+\phi_{0-})\nonumber\\
c_{xy}^{+-} & = & -(\phi_{+-}+\phi_{00})-(\phi_{+0}+\phi_{0-})\nonumber.
\end{eqnarray}
Then from $H=c_{xy}^{++}-c_{xy}^{-+}+c_{xy}^{--}-c_{xy}^{+-}$ it follows that 
\begin{eqnarray}\label{eqn:H_lin}
H
&=&\;\;\phi_{-+}-2\,\phi_{0+}\;\,+\phi_{++}\\
& &\!\!\!\!\!-2\,\phi_{-0}\,+4\,\phi_{00}-2\,\phi_{+0}\nonumber\\
& &\!\!+\,\phi_{--}-2\,\phi_{0-}\;+\phi_{+-}.\nonumber
\end{eqnarray}

\noindent\textbf{For trilinear splines}, label the eight cells surrounding a junction point (taken temporarily to be the origin) by indexes $\alpha = \textrm{sgn}(x)$, $\beta=\textrm{sgn}(y)$ and $\gamma=\textrm{sgn}
(z)$. Then
\begin{equation}
\phi^{\alpha,\beta,\gamma}(x,y,z)=\phi_{0}+c_x^\alpha x+c_y^\beta y + c_z^\gamma z+c_{xy}^{\alpha,\beta}x y  +c_{yz}^{\beta,\gamma}y z  +c_{zx}^{\gamma,\alpha}z x +c_{xyz}^{\alpha,\beta,\gamma}x y z,
\end{equation}
and solving for $c_{xyz}^{\alpha,\beta,\gamma}$ in terms of the values of $\phi$ at the junctions gives
\begin{eqnarray}
a^3\alpha\beta \gamma\,c_{xyz}^{\alpha,\beta,\gamma}&&= -\phi_0
+\phi(\alpha a,0,0)+\phi(0,\beta a,0)+\phi(0,0,\gamma a) \\
&&-[\phi(\alpha a,\beta a,0)+\phi(0,\beta a,\gamma a)+\phi(\alpha a,0,\gamma a) ]+\phi(\alpha a,\beta a,\gamma a). \nonumber
\end{eqnarray}
Then, similarly to the 2D case, $h_{i,j,k}$ at the centre is
\begin{eqnarray}
 && h_0 =-a^3\sum\nolimits_{\alpha \beta \gamma}\alpha \beta \gamma\,c_{xyz}^{\alpha,\beta,\gamma}= 8\phi_0 -\sum\nolimits_{\alpha \beta \gamma}\phi(\alpha a,\beta a,\gamma a)\\
&&-4\,\big[\sum\nolimits_{\alpha}\phi(\alpha a,0,0)+\sum\nolimits_{\beta}\phi(0,\beta a,0)+\sum\nolimits_{\gamma}\phi(0,0,\gamma a)\big] \nonumber\\
&&+2\,\big[\sum\nolimits_{\alpha \beta}\phi(\alpha a,\beta a,0)+\!\sum\nolimits_{\beta \gamma}\phi(0,\beta a,\gamma a)+\!\sum\nolimits_{\alpha \gamma}\phi(\alpha a,0,\gamma a) \big].\nonumber
\end{eqnarray}
The simple result is that $h_{ijk}$ is found by adding the weighted values of $\phi$ at the 27 junctions in the surrounding cube, with weight 8 at the centre, -4 for the 6 nearest neighbours, +2 for the 12 next nearest, and -1 for the 8 furthest points.

\vspace{2mm}
\noindent\textbf{Biquadratic splines}. Given the set of 1D splines, we have $\phi, \phi_x$ and $\phi_y$ at each junction point. 
%For each cell we have eight independent quantities:
%$\phi(i,j)$, $\phi(i+1,j)$, $\phi(i,j+1)$, $\phi(i+1,j+1)$, $\phi_x(i,j)$, $\phi_y(i,j)$, $\phi_x(i,j+1)$, $\phi_y(i+1,j)$. These eight quantities, however do not necessarily completely define a biquadratic for that cell, which generally requires nine quantities, such as $\phi$, $\phi_x$, $\phi_y$, $\phi_{xx}$, $\phi_{xy}$, $\phi_{yy}$, $\phi_{xxy}$, $\phi_{xyy}$, $\phi_{xxyy}$ at one junction. If it is necessary, it may be useful to use $\phi_{xy}$ to move from one cell to a neighbor, because this quantity must be the same in all four cells that surround a junction where $\phi_x$ and $\phi_y$ are continuous.
It is possible to find all partial derivatives at all junctions of a biquadratic spline, given the set of 1D splines on the gridlines (and in some cases one extra quantity, such as the value of $\phi_{xy}$ at one junction). The spline in each cell can be expressed in terms of its derivatives at one corner and these nine independent constants can be solved for in terms of the known values of $\phi$, $\phi_{x}$ and $\phi_{y}$ at the junctions, together with $\phi_{xy}$ at one junction. The derivatives at each junction can then be easily calculated. By this process, one can continue through all the cells of the spline; but inconsistencies may arise, indicating that no regular biquadratic spline is consistent with the initial assumptions.

To determine the evolvable form, we need only $\phi_{xxyy}$ at the internal junctions [to find $h(i,j)$], but on the boundary we also need $\phi_{xxy}$ and $\phi_{xyy}$, and at the corners $\phi_{xy}$.

\vspace{2mm}
\noindent\textbf{When the first-derivatives in a biquadratic spline are not continuous}, we require terms in $f_1(x-a_i)$ and $f_1(y-b_j)$. Then we write
\begin{eqnarray}\label{eqn:phi_hbc}
\phi(x,y)=\sum\nolimits_{i,j}&&\big{[}h_{i,j}f_2(x-a_i)\,f_2(y-b_j)+k_{i,j}f_1(x-a_i)\,f_1(y-b_j)\\
&&+c^x_{i,j}f_1(x-a_i)\,f_2(y-b_j)+c^y_{i,j}f_1(y-b_j)\,f_2(x-a_i)\big{]}\nonumber.
\end{eqnarray}
The coefficients $h_{i,j}$, $k_{i,j}$, $c^x_{i,j}$, $c^y_{i,j}$ can be determined from the derivatives of $\phi$ at the junction $i, j$ as follows. From Eq.(\ref{eqn:phi_hbc}), $\phi_{x,y}$ contains a term $k_{i,j}f_0(x-a_i)\,f_0(y-b_j)$ and there are no other terms implying a step in both $x-a_i$ and $y-b_j$. Therefore, 
\begin{equation}
k_{i,j}=\phi^{++}_{x,y}+\phi^{--}_{x,y}-\phi^{+-}_{x,y}-\phi^{-+}_{x,y},
\end{equation}
where $\phi^{++}_{x,y}=\phi_{x,y}(a_i+\epsilon,b_j+\epsilon)$ with $\epsilon>0$ and small, and similarly for $\phi^{--}_{x,y},\,\phi^{+-}_{x,y}$ and $\phi^{-+}_{x,y}$. For a regular spline $\phi_{x,y}$ does not change when crossing from one cell to another (at a junction) and therefore $k_{i,j}$ can be nonzero only at corners of the boundary. 

In the same way, we can determine $c^x_{i,j}$ from $\phi_{xyy}$, $c^y_{i,j}$ from $\phi_{xxy}$ and $h_{i,j}$ from $\phi_{xxyy}$. For a regular spline, $c^x_{i,j}$ and $c^y_{i,j}$ can be non-zero only on the boundary.

\vspace{2mm}
\noindent\textbf{Example: the spline in Fig.\,\ref{fig:BiQ}}.   
The explicit unique symmetric 2D spline is (using the cell labels in Fig.\,{\ref{fig:data1}):
\begin{eqnarray}
&&R1:  x^2(1+y)^2 \\
&&R2: (2-x)(3x-2)(1+y)^2 \nonumber \\
&&R3: \frac{1}{4}(x-2) (8-12x+16y-24xy-26y^2+31xy^2) \nonumber \\
&&C1: x^2+y^2+2(x^2y+xy^2)-\frac{17}{4}x^2y^2 \nonumber \\
&&C2: \frac{1}{4}(2-x)(y-2)[36-38(x+y)+33xy]. \nonumber
\end{eqnarray}
For the cells $L1, L2$ and $L3$, interchange $x$ and $y$ in the expressions for cells $R1, R2$ and $R3$.

The derivatives used for evolution are shown in Fig.\,{\ref{fig:data1}. Due to the symmetry, $\phi_{xy}(i,j)=\phi_{xy}(j,i)$ and $\phi_{xxy}(i,j)=\phi_{xyy}(j,i)$. From these derivatives, the coefficients $h(i,j)$ and $c(i,j)$ in Eq.(\ref{eqn:phi_hbc}) are easily calculated and are shown in Fig.\,{\ref{fig:data2}. There is only one non-zero value for $k(i,j)$: $k(4,4)=\phi_{xy}(4,4)=-4$. Therefore the evolvable form of the spline is
%\begin{eqnarray}\label{eqn:biQev}
%\phi(x,y)=\sum\nolimits_{i,j}&&h_{i,j}f_2(x-a_i)\,f_2(y-b_j)+\nonumber\\
%&&b_{4,4}f_1(x-a_4)\,f_1(y-b_4)+\nonumber\\&&\sum_{i=1}^4c_{i}{\big[}f_1(x-a_i)\,f_2(y-b_j)+\nonumber\\&&f_1(y-b_j)\,f_2(x-a_i){\big[}.
%\end{eqnarray}
\begin{eqnarray}\label{eqn:biQev}
\phi(x,y)=\sum\nolimits_{i,j=1}^4&&h_{i,j}f_2(x-a_i)\,f_2(y-a_j)+
k_{4,4}f_1(x-a_4)\,f_1(y-a_4)\\ &&+\sum\nolimits_{i=1}^4 c_{i}{\big[}f_1(x-a_4)\,f_2(y-a_i)+f_2(x-a_i)\,f_1(y-a_4){\big]},\nonumber
\end{eqnarray}
where $\mathfrak{a}=(-1,0,1,2)$ and $\mathfrak{c}=(8,-26,32,-14)$.

\begin{figure}[h!]
\centering
\includegraphics[width=10cm]{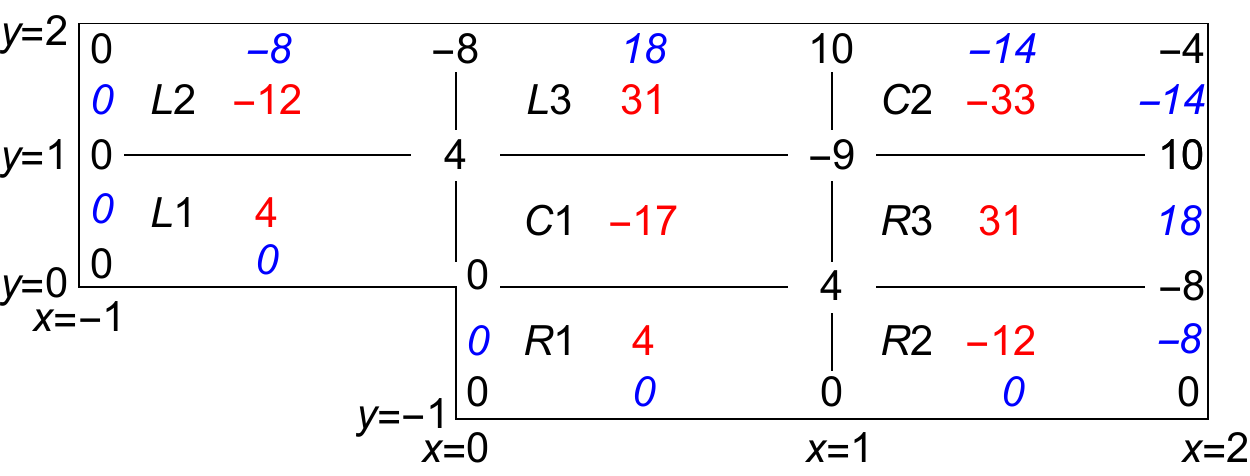}
\caption{The derivatives used to evolve the spline shown in Fig.\,\ref{fig:BiQ}. The numbers at the junctions are the values of $\phi_{xy}$ and the italicised numbers inside (and at the centre of) the bounding edge of each outer cell give $\phi_{xxy}$ on the horizontal lines or $\phi_{xyy}$ on the verticals, while the values of $\phi_{xxyy}$ lie at the centre of each cell. The labels $L1$, $L2$, ... label each cell.}
\label{fig:data1}
\end{figure}

\begin{figure}[h!]
\centering
\includegraphics[width=10cm]{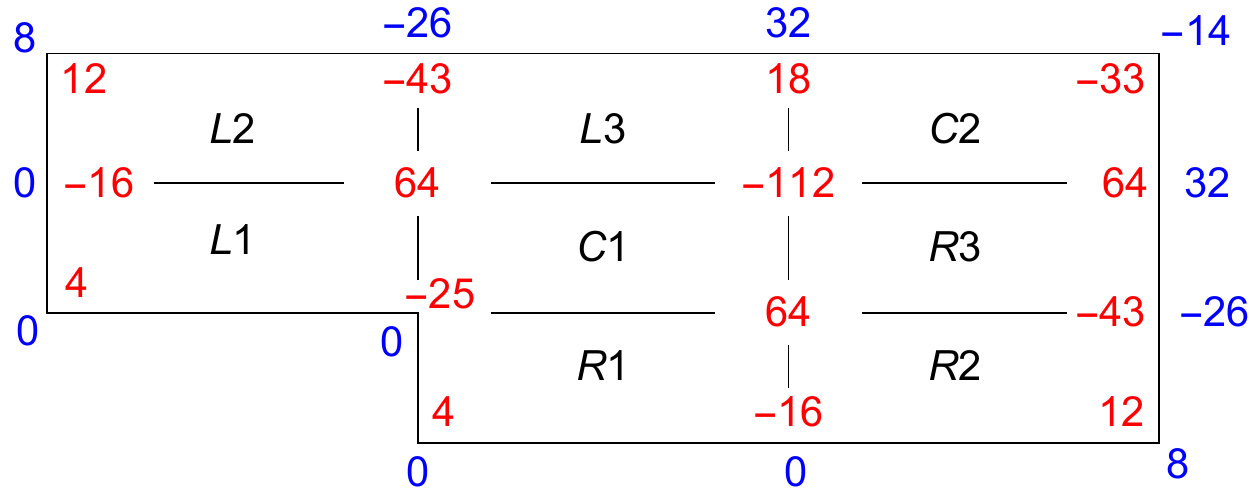}
\caption{The quantities calculated from the derivatives in Fig.\,\ref{fig:data1} and used to evolve the spline shown in Fig.\,\ref{fig:BiQ}. The number at each junction (and inside the boundaries) gives $h(i,j)$, calculated from $\phi_{xxyy}$ in the four neighbouring cells. The numbers outside the boundary give $c(i,j)$, calculated from the two neighbouring values of $\phi_{xxy}$ (or $\phi_{xyy}$).}
\label{fig:data2}
\end{figure}
\pagebreak

\end{document}